\documentclass[aps,prb,reprint,amsmath,superscriptaddress,showpacs]{revtex4-1}
\usepackage{graphicx}

\newcommand{\siesta}{\textsc{siesta}}
\newcommand{\transiesta}{\textsc{tran\-siesta}}

\newcommand{\energy}{\ensuremath{\omega}}

\begin{document}

\begin{abstract}
  We have recently shown that by using a scaling approach for randomly
  distributed topological defects in graphene, reliable estimates for
  transmission properties of macroscopic samples can be calculated
  based even on single-defect calculations [A. Uppstu {\it et al.},
  Phys. Rev. B {\bf 85}, 041401 (2012)].  We now extend this approach
  of energy-dependent scattering cross sections to the case of
  adsorbates on graphene by studying hydrogen and carbon adatoms as
  well as epoxide and hydroxyl groups.  We show that a qualitative
  understanding of resonant scattering can be gained through density
  functional theory  results for a single-defect system, providing a
  transmission ``fingerprint'' characterizing each adsorbate  type. 
   This information can be used to reliably predict the elastic mean free path 
  for moderate defect densities directly using {\it ab-initio} methods.
  We present tight-binding parameters for carbon and epoxide adsorbates, obtained to match the
  density-functional theory based scattering cross sections.
\end{abstract}


\title{Ab-initio transport fingerprints for resonant scattering in graphene}

\author{Karri Saloriutta}
\email{karri.saloriutta@aalto.fi}
\affiliation{COMP Centre of Excellence,
  Department of Applied Physics, Aalto University School of Science,
  P.O. Box 11100, FIN-00076 AALTO, Finland}

\author{Andreas Uppstu}
\author{Ari Harju}
\affiliation{COMP Centre of Excellence and Helsinki Institute of Physics,
  Department of Applied Physics, Aalto University School of Science,
  P.O. Box 14100, FIN-00076 AALTO, Finland}

\author{Martti J. Puska}
\affiliation{COMP Centre of Excellence,
  Department of Applied Physics, Aalto University School of Science,
  P.O. Box 11100, FIN-00076 AALTO, Finland}

\date{\today}

\maketitle

\section{Introduction}

Electron transport in graphene has been studied widely from different
angles due to its fundamental and application-derived importance (for
reviews, see Refs. \onlinecite{CastroNeto2009} and
\onlinecite{DasSarma2011}).  Depending on the nature of the graphene
sample and the substrate on which transport is measured, different
scattering mechanisms have been proposed.  For graphene on SiO$_{2}$,
scattering from charges in the substrate is believed to be an
important contribution.\cite{Zhang2009f, Adam2007b} On the other hand,
resonant scattering from impurities, such as adsorbates and defects, can
be large on a high-$\kappa$ substrate and depend on the way the
sample has been synthesized and cleaned.\cite{Giannazzo2011,
  Ferreira2011} We concentrate here on this latter case of resonant
scattering.  In addition to its importance for understanding the
fundamental properties of experimental samples, which are seldom fully
without adsorbed species, understanding resonant scattering might be
relevant for the purposeful modification of graphene properties
through functionalization.

Resonant scattering in graphene nanostructures is often studied
numerically by a tight-binding model (TB) so that numerous
calculations are performed for samples containing a set number of
randomly placed defects.  The results are then averaged in order to
obtain for every defect concentration defect-specific properties, such
as the elastic mean free path or the conductance, as a function of the
charge carrier energy. This kind of ensemble averaging is a costly
procedure even in the lightweight TB picture of the electronic
structure. Therefore, based on prior work on electron transmission in
silicon nanowires \cite{Markussen2007,Markussen2010}, we have recently
applied the concept of energy-dependent scattering cross section to
scale the transport properties of single defects to those of
macroscopic samples with homogeneous defect
concentrations.\cite{Uppstu2012}

The TB calculations can be parametrized based on the density
functional theory (DFT) results or empirically based on experimental
data. However, the modeling of defects beyond simple alterations in
the topology, i.e. the modeling of impurities, chemical adsorbates or
even carbon adatoms that create different bond configurations, becomes
challenging since the parametrization should catch all the features of
the chemical bonds relevant for the transport properties.  A
systematic determination of an adequate TB model is thus difficult and
there are competing parametrization even for simple cases such as
hydrogen.\cite{Wehling2010c, Robinson2008} Therefore, in this work we
study the determination of the effective scattering cross sections
directly from the DFT results for systems containing only a single
defect. We also show that the single defect scattering cross section
can be used to determine the optimal TB model parameters for transport
calculations.  In this way we fit the TB parameters using a quantity
directly related to the resonant electron scattering by the defect in
question.

First, we study hydrogen atoms and hydroxyl 
groups which are adsorbed on 
top of a carbon atom. The hydrogen atom is a good test case
because there are different TB models to describe it and we can
compare results given by them with DFT results.  The adsorption
of the hydroxyl group increases the complexity from that of the
hydrogen atom and the DFT results for the scattering cross section show
clear 
deviation from the hydrogen case. 
Oxygen is adsorbed in
the form of epoxide on the carbon-carbon bridge which also increases
the complexity of the chemical bonding. We compare the DFT results
with those of a recent TB parametrization \cite{Leconte2011a, Cresti2011} and
find out that 
the scattering cross sections from these two methods match
apart from some details in regions of rapid variations. We show that these details can be improved by optimizing the TB parameters.
 
Furthermore, we present a TB model for a carbon adatom bound to the
carbon-carbon bridge. The parameters are obtained by optimizing the TB
scattering cross section to match the corresponding DFT prediction.

All the adsorbates we study are important from the experimental point
of view and have potential for technological applications. Hydrogen is
an important functionalizer that turns graphene into \emph{graphane},
a band insulator, at full coverage.\cite{Sofo2007, Elias2009} At lower
coverage, evidence for metal-insulator transition has been
seen.\cite{Bostwick2009} Oxygen can break graphene sheets or unzip
carbon nanotubes to GNRs.\cite{Xu2010, Sun2011} Epoxide is also a
component of graphene oxide,\cite{Boukhvalov2008e, Yan2009a} which has
been used for graphene synthesis\cite{Park2009b} and sensor
applications.\cite{Robinson2008, Lu2009a} Carbon adatoms, on the other
hand, are created during irradiation of graphene, e.g., when
intentionally creating vacancy and bond rotation
defects.\cite{Banhart2010}

The organization of the paper is as follows. In Sec.\ II, we describe
the electronic structure models and methods used and review the main
ideas for the the scattering cross section concept. In Sec.\ III, we
study the effect of different approximations on the scattering
cross section for the atop-adsorbed hydrogen and hydroxyl as test
cases. In Sec.\ IV, we concentrate on the effect of the adsorbate
chemistry by presenting results for oxygen and adatom carbon
preferring bridge-site adsorption.

\section{Method}

\begin{figure}
  \centering
  \includegraphics[width=\columnwidth]{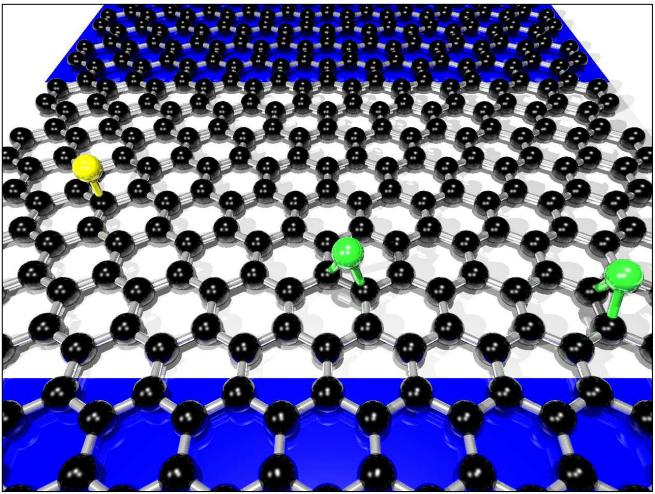}
  \caption{\label{fig:system} (Color online) Two-probe transport
    geometry consisting of two ideal leads and a central scattering
    region.  The graphene sheet is oriented with the transport along
    the armchair direction. Upper and lower shadowed (blue/gray)
    regions indicate the leads. The left-most adsorbate on the
    scattering region is on top of a carbon atom and the others show
    two of the three bridge positions with different orientations with
    respect to the current.}
\end{figure}

Figure~\ref{fig:system} shows the schematic setup for a two-probe
transport calculation. The current flows between the shaded leads.
The graphene sheet is oriented so that the armchair
edge is parallel to the transport direction.  The ribbons we study are
thus armchair graphene nanoribbons (AGNR), and we restrict our DFT
calculations to shown ribbons with 23 carbon dimer lines 
(23-AGNR).

The transmission through the ribbon or sheet is affected by the
scattering of electrons off adsorbates and studied by the ``standard''
setup for ab-initio transport calculations, i.e., the combination of
DFT and nonequilibrium Green's functions (NEGF).\cite{Haug2008} We use
the \transiesta\ implementation, a part of the \siesta\ DFT
package.\cite{Soler2002, Brandbyge2002}

We compare the full ab-initio approach with the widely-used
nearest-neighbor TB model of graphene.  The TB approach is based on
the Hamiltonian
\begin{equation}
  \label{eq:tb_hamiltonian}
  H = \sum_{<i,j>} t c_{i}^{\dagger}c_{j} + h.c.,
\end{equation}
where the sum is taken over the nearest neighbor pairs and the nearest
neighbor hopping element $t$ is $-2.7$ eV.  An ideal (unrelaxed)
vacancy can be described simply by removing the hopping elements to
a particular carbon site.  This is also the simplest model for a
hydrogen adatom on top of a carbon atom since it describes the removal
of the carbon $\pi$-bond due to the creation of a
carbon-hydrogen bond.

In order to more realistically describe an adatom on top of a carbon
atom, the simple TB model can be complemented by an Anderson
impurity level added to the Hamiltonian in \eqref{eq:tb_hamiltonian}, i.e.,
\begin{equation}
  \label{eq:tb_impurity}
  H = \sum_{<i,j>} \left ( t c_{i}^{\dagger} c_{j} +  t c_{j}^{\dagger} c_{i} \right ) + e_A d^{\dagger}d + t_A
  d^{\dagger}c_{m} + t_A c^{\dagger}_{m} d ,
\end{equation}
where the carbon atom with the index $m$ is coupled with hopping $t_A$
to the impurity level with energy $e_A$. This simple and versatile
model has been widely used to model hydrogen and various other
adsorbates, such as the hydroxyl group and hydrocarbons, that bind on
top of a carbon atom.~\cite{Robinson2008,Wehling2010c, Yuan2010} For
hydrogen, the parameters giving results corresponding well to DFT
calculations\cite{Wehling2010c} are $t_A = -2t$\ and $e_A=t/16$\,
though other values have also been proposed.\cite{Robinson2008}

In our DFT calculations, we use the double $\zeta$ polarization (DZP)
basis set for relaxing the structures with the Broyden algorithm so
that ionic forces are less than 0.04 eV/\AA.  We use the PBE
functional for exchange and correlation.\cite{Perdew1996} The basis
set is automatically constructed by \siesta\ based on an energy shift
of 80 meV and a split norm of 0.3 for all elements apart from hydrogen
for which 0.5 is used.  The grid interval corresponding to the Fourier
cutoff energy of 250 Ry is used for the real-space grid. The periodic
images are separated by 19 \AA\ of vacuum.  These parameters lead to a
graphene lattice constant of 1.43 \AA\ which is used in all the
calculations.  For transport the basis set is decreased to the single
$\zeta$ polarization (SZP) and the grid to 150 Ry.

In addition to nanoribbons, we study the transmission through periodic
systems, i.e., graphene sheets.  The sheets are oriented so that the
transport is along the armchair direction.  The system sizes are
chosen so that the scattering cross section defined below is
well-converged requiring the supercells to correspond to nanoribbons
from 24-AGNR to 32-AGNR depending on the adsorbate studied.  The
periodicity is handled by a Monkhorst-Pack k-point sampling
perpendicular to the current direction, with 3 to 6 k-points used for
relaxation and 21 to 45 for transport.

Following Refs. \onlinecite{Uppstu2012} and
\onlinecite{Markussen2010}, we define the scattering cross section:
\begin{equation}
  \sigma(\energy) = W \frac{T_0(\energy) - \left<T(\energy)\right>}{\left<T(\energy)\right>},
  \label{eq:sigma}
\end{equation}
where $W$ is the lateral width of the supercell for periodic graphene
or the width of the nanoribbon for GNRs.  $T_0(\energy)$ is the
transmission through the pristine system without a defect and
$\left<T(\energy)\right>$ is the transmission through the defect
averaged over the defect locations and orientations.  The scattering
cross section is a rather abstract quantity.  Based on it, we can
however produce estimates for the transmission through a defective
sample with some finite defect density.  For a sample of length $L$
with different defect species with scattering cross sections
$\sigma_i$ and densities $n_i$ we get
\begin{equation}
  \label{eq:t_ave}
  \left< T(\energy)\right> = \frac{T_0(\energy)}{1 + L\sum_i n_i \sigma_i(\energy)},
\end{equation}
within the Ohmic regime. We can also estimate quantities that can be
directly experimentally measured.  The elastic mean free path, for
example, is given by
\begin{equation}
  l_e(\energy) = \sum_i \frac{1}{n_i \sigma_i(\energy)}.
  \label{eq:elastic}
\end{equation}

Both equation \eqref{eq:sigma} and \eqref{eq:t_ave} are defined within
the Ohmic regime where the transmissions can be averaged directly as
the transmission has a Gaussian probability distribution.  The cross
section can, however, also be defined within the localized regime, as
was done in our previous work.\cite{Uppstu2012} This is seldom needed
for cross sections based on single defect calculations and we
concentrate on the Ohmic regime here.
 
One weakness of using the scaling approach in this manner is that we
are assuming a random distribution of absorbents on graphene.  If this
is not the case and the disorder is \emph{correlated}, the approach
loses some of its validity.  Indeed, there are indications that oxygen
adsorbates, for example, tend to favor clustering since this is
preferred energetically.~\cite{Yan2009a} Correlation between
\emph{Coulombic} scatterers, on the other hand, can be used to explain
the transport properties of graphene samples with no resonant
scattering.~\cite{Li2011c}

DFT can be used to study spin-polarized transport and thus could be
used to calculate a spin-dependent scattering cross section.  All our
calculations, however, are done spin-compensated despite the fact that
some of the systems we study in fact have a finite magnetic moment in
the ground state.  The addition of spin-dependence would make the
analysis of disordered systems much more complicated since long-range
ordering of the spins is likely.\cite{Soriano2011a, Leconte2011b} We
also consider just armchair ribbons since zigzag ribbons have spin
polarized edges.  Adding spin-polarization to the calculations would
split the resonance peaks at the Fermi energy into two closely spaced
spin-polarized peaks.\cite{Yazyev2007} The scattering cross section
for the both spin-channels would then be very similar to the
spin-compensated cross sections presented here.

\section{Atop adsorbates: the atomic hydrogen and hydroxyl group}

To test the sensitivity of determining the adsorbate scattering cross
sections on electronic-structure models and nanoribbon or bulk
graphene geometries used, we first consider perhaps the simplest
conceivable case, the hydrogen atom. This system is convenient since,
as shown below, the TB model is enough to capture all the relevant
physics with limited computational resources and even very large
system sizes directly relevant to experiments can be studied.

\begin{figure}[h]
  \centering
  \includegraphics[width=\columnwidth]{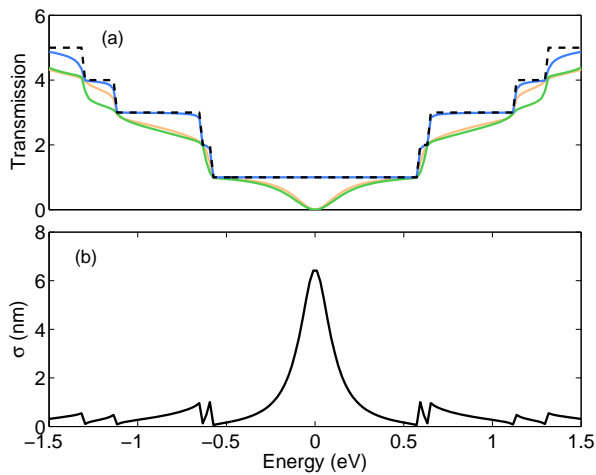}
  \caption{\label{fig:vac_a23} (Color online) (a) Ideal electron transmission for a
    pristine 23-AGNR (dashed line) and transmissions for a 23-AGNR with
    one hydrogen atom adsorbed on three different locations (solid
    lines).  (b) Scattering cross sections calculated by averaging
    over all the different adsorbate locations across the ribbon.  The
    vacancy TB model is used. Here and in all the figures below, the
    Fermi energy is taken to be zero.}
\end{figure}

As discussed above, the simplest TB model for hydrogen on graphene
treats it as a vacancy.  This is actually quite a poor model for a
real vacancy since it does not account for the relaxation caused by
the reorganization of the dangling bonds.\cite{Banhart2010}
Figure~\ref{fig:vac_a23} (a) shows the electron transmissions for a
pristine 23-AGNR and for a 23-AGNR with an adsorbed hydrogen in three
different locations calculated with the vacancy TB model. Two of the
three latter transmissions are very similar, with a dip right in the
middle of the first transmission plateau at the Fermi energy while for
one adsorbate location there is no dip.  This is a general feature for
metallic armchair nanoribbons: when hydrogen is adsorbed on a dimer
indexable by 3N, backscattering is suppressed due to the electronic
structure of the ribbon.\cite{Deretzis2010, Choe2010} Therefore the
results for AGNRs shown below are calculated with Eq.~(\ref{eq:sigma})
by averaging over all different adsorption sites.  As an example of
this, Fig.~\ref{fig:vac_a23} (b) shows the scattering cross section
for the hydrogen adsorbate.  The clearest feature is the strong peak
at the Fermi energy that is caused by a resonant level at that energy
and also shows up in the density of states.  The other characteristic
feature is the spikes further away from the Fermi level. They are due
to the density of states van Hove singularities caused by the band
edges of the quasi one-dimensional system.

\begin{figure}
  \centering
  \includegraphics[width=\columnwidth]{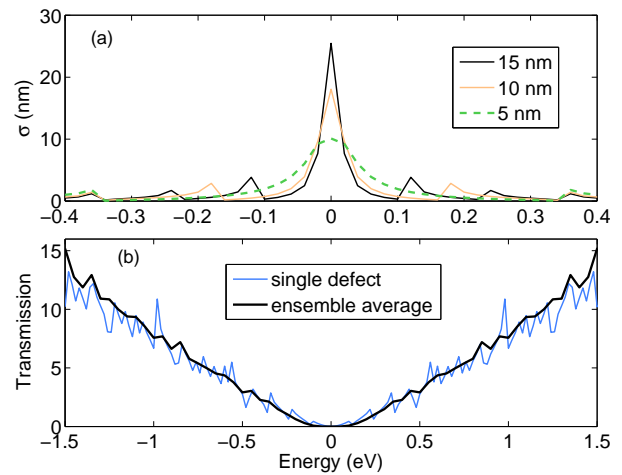}
  \caption{\label{fig:vac_size} (Color online) (a) Scattering cross
    sections for a hydrogen atom on AGNRs with different widths.  The
    vacancy TB model is used. (b) Transmission through a 100 nm long
    125-AGNR with 0.1 \% vacancy defect density.  Comparison between
    results obtained by averaging over an ensemble of 50 samples with
    random defect locations (thick line) and an estimate from a single
    defect scattering cross section calculation (thin line). Notice
    the different energy scales in the two panels.}
\end{figure}

Ideally, we would like to go from the scattering cross section of a
narrow ribbon calculated using a full DFT approach by scaling to wider
and longer systems using Eq. \eqref{eq:t_ave}.  Unfortunately, the
details of the scattering cross section depend on the width of the
nanoribbon. This is shown in Fig. \ref{fig:vac_size} (a) for the case
of a hydrogen atom adsorbed on AGNRs of different width.  The
vacancy-model TB results indicate that the actual positions of the van
Hove peaks depend on the ribbon width and that the central peak
becomes thinner and higher as the ribbon width increases.  It is
however clear that it is possible to get a good qualitative
understanding of the transmission for wider ribbons based on the
scattering cross section calculated for a narrow ribbon.

Figure \ref{fig:vac_size} (b) shows the use of Eq. \eqref{eq:t_ave} by
comparing an estimate based on the scattering cross section from a
single defect calculation to an averaged transmission through a 100 nm
long 125-AGNR.  As long as the transmission is large enough, a very
good estimate for the transmission through the longer sample can be
generated.  Close to the Fermi energy, the transmission is within the
localized regime and the estimate based on Ohmic transmission
(Eq. \ref{eq:t_ave}) is larger than the average.  Although here we
concentrate on the Ohmic regime, the localized regime can also be
treated within the scattering cross section approach as shown in our
previous work.\cite{Uppstu2012}

\begin{figure}
  \centering
  \includegraphics[width=\columnwidth]{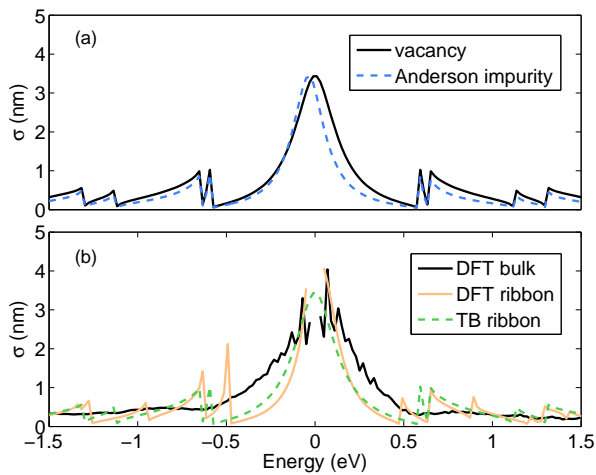}
  \caption{\label{fig:hydrogen} (Color online) (a) Scattering cross
    section for a hydrogen atom on a 23-AGNR.  The results of the
    Anderson impurity hydrogen model with $t_A = -2t$\ and
    $e_A=t/16$\ and the vacancy-model are shown. 
    (b) Scattering cross section for a hydrogen atom on a bulk geometry from DFT and
    a 23-AGNR calculated with both the DFT and TB models. 
    }
\end{figure}

The vacancy TB model does not directly correspond to a real
monovacancy or a hydrogen adsorbate. For comparison we have used
also the impurity-model of Eq. \eqref{eq:tb_impurity}. Figure
\ref{fig:hydrogen} (a) shows the calculated scattering cross sections
for a 23-AGNR with one adsorbed hydrogen atom. In this case, the
cross section from the impurity Hamiltonian is quite close to that
from the vacancy-model.  The impurity-model results in a slight
asymmetry and shift with respect to the vacancy-model. Thus, the
simpler vacancy-model seems to give a quite adequate scattering cross
section for the adsorbed hydrogen.

The great advantage of the scattering cross section formalism is that
it makes possible to study scattering in a fully ab-initio manner,
based directly on DFT transport calculations.  However, dealing with
graphene nanoribbons requires still a remarkable amount of
computational time because all the different adsorbate locations
across the ribbon need to be solved self-consistently including the
lattice relaxation prior to the transport calculation. Moreover, the
separate DFT calculations are limited to rather small ribbons,
typically below 5 nm in width.  We can, however, significantly lower
the computational cost by studying the scattering cross sections of
bulk systems, that is, systems with periodic boundary conditions
perpendicular to the transport direction.

Figure \ref{fig:hydrogen} (b) shows the scattering cross section for a
hydrogen atom on a 23-AGNR calculated both within the DFT and TB
schemes.  The quite close correspondence indicates that the simple TB
model is quite suitable for simple adsorbates, such as hydrogen, that
bind on top of a carbon atom. The cross section for a hydrogen atom in
bulk graphene, obtained by imposing periodic boundary conditions as
described above, is also shown. In the bulk geometry, the van Hove
singularities are smoothed out and the central peak is correspondingly
broadened.  The same considerations thus apply here as for the vacancy
hydrogen model discussion relating to Fig. \ref{fig:vac_size}: a good
qualitative understanding is possible based on a bulk DFT scattering
cross section but some of the details are necessarily lost when
translating from the infinite bulk system to a ribbon.

\begin{figure}
  \centering
  \includegraphics[width=\columnwidth]{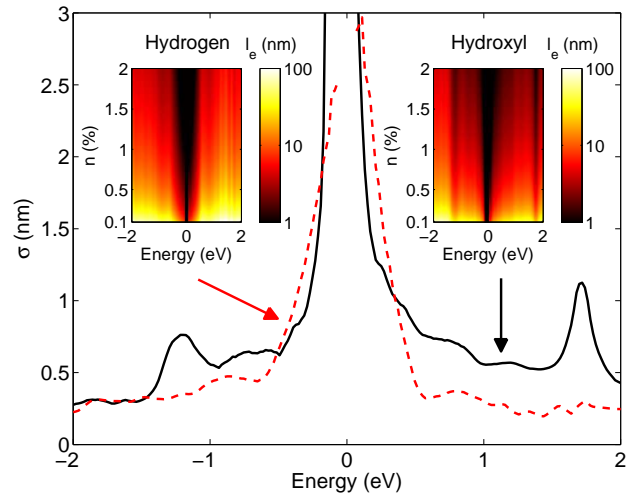}  
  \caption{(Color online) Scattering cross section for a hydrogen atom (dashed line)
    and a hydroxyl group (solid line) adsorbed on the bulk graphene
    calculated with DFT. The insets show the corresponding elastic
    mean free paths 
    for different adatom densities. The scales of the insets have been cut at 1 nm and 100 nm.}
  \label{fig:oh_ga}
\end{figure}

As an example of atop adsorption in which the details of the chemical
bonding become important and the vacancy TB model cannot predict the
scattering cross section accurately enough, we consider the hydroxyl
(OH) group. We perform a DFT bulk calculation for the adsorbed OH and
compare the cross section with that for adsorbed hydrogen. Figure
\ref{fig:oh_ga} shows that although both adsorbates create a prominent
resonance close to the Dirac point, hydroxyl adsorption results in a
more asymmetric central peak and additional smaller resonances farther
away from the Dirac point.  These features serve as the
distinguishable fingerprints between hydrogen and hydroxyl.  This
shows the usefulness of the scattering cross section as a scattering
fingerprint, the bulk cross section especially depends only on the
adsorbate species and has a characteristic shape for each adsorbate.
The insets in Fig. \ref{fig:oh_ga} show the estimated elastic
mean free paths for the hydrogen and hydroxyl adsorbates, based on the
single-defect cross sections. Even small concentrations of such
adsorbates lead to very short mean free paths close to the Dirac
point, and the effects of the side peaks in the scattering cross
section can also be clearly seen.

\section{Bridge adsorbates: epoxide and carbon adatom}

We now turn to the more complicated case of bridge-bonded adsorbates.
As previously, when defect-defect interactions are negligible, a
single-defect DFT calculation may be used to estimate the mean free
path of a large-scale system. However, with large concentrations of
defects, the scatterers start to interact, and to capture these
effects large-scale calculations are needed. These are most
conveniently done using the TB method, which is able to simulate
systems with large numbers of adsorbates. The TB model parameters can
be optimized against the corresponding DFT-based single-defect
scattering cross section. Here we present optimized TB parameters for
epoxide and carbon adsorbates.

\begin{figure}
  \centering
  \includegraphics[width=\columnwidth]{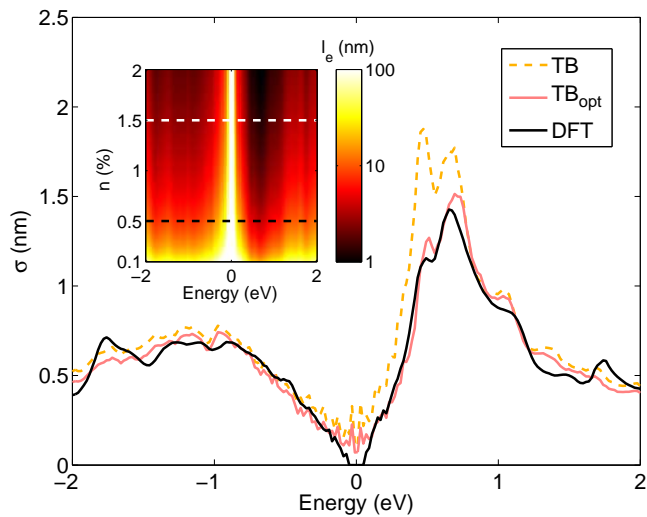}
  \caption{\label{fig:epoxide} (Color online)
    Scattering cross section for an
    oxygen adatom on bulk graphene (epoxide) averaged over the three
    different bridge sites. The results of DFT
    and two TB approaches with different parameter sets are shown. 
    The inset shows the elastic mean free path 
    for different adsorbate concentrations. The scale of the inset has been cut at 1 nm and 100 nm.}
\end{figure}

\begin{figure}
  \centering
  \includegraphics[width=\columnwidth]{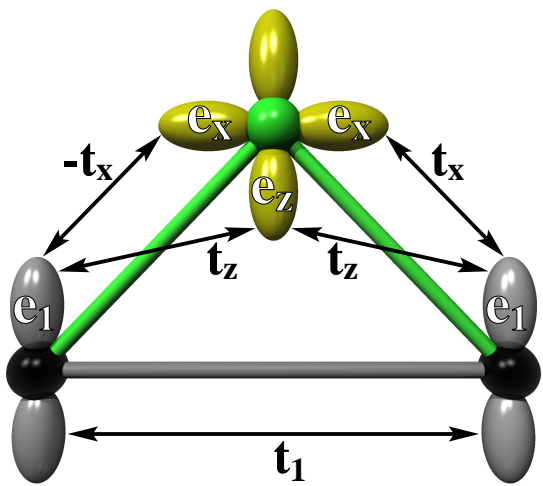}
  \caption{\label{fig:hops} (Color online) Tight-binding hopping terms
    $t$ between the two states of the bridge adatom (top) and the two
    carbon atoms on the graphene plane. The state energies $e$ are
    also shown.}
\end{figure}

Oxygen preferably binds to graphene on the bridge site between carbon
atoms and forms an epoxide group.  Figure \ref{fig:epoxide} shows the
scattering cross section calculated for epoxide on bulk graphene.
Unlike the case of hydrogen, there are now three different bridge
sites with respect to electron current direction and one needs to
average the scattering cross section over them.  The most prominent
feature of the scattering cross section is the peak above the Fermi
energy, corresponding to the oxygen acceptor state in the density of
states.  This quasi-localized state couples with the electronic states
of graphene causing increased scattering for electrons as compared to
holes.  This is also reproduced by two different sets of TB results,
also presented in Fig.~\ref{fig:epoxide}.  The yellow (light grey)
curve has been obtained by using the parameters from
Ref.~\onlinecite{Leconte2010}, but the agreement between TB and DFT
can be improved by optimizing the parameters to better match the
scattering cross section. The optimized parameter set is $e_1=-3.79$
eV, $e_x=-2.78$ eV, $e_z=0.04$ eV, $t_1=-1.15 \, t$, $t_z=-1.22 \, t$,
and $t_x=2.50 \, t$ (for notation, see Fig.~\ref{fig:hops}).  In
agreement with Ref.~\onlinecite{Leconte2010}, we have used $t=-2.6$
eV.  The resulting cross section is shown by the red (dark grey) curve
in Fig.~\ref{fig:epoxide}. As can be seen, the main resonance is
captured more accurately by the optimized model. However, additional
small resonances close to $\pm 2$ eV are not reproduced. These are
better seen in the inset, which shows the DFT-based elastic mean free
path $l_e$ for different concentrations of epoxide
adsorbates. Away from the Dirac point, the most prominent feature is
located roughly at 0.6 eV, where the resonance produces a clear drop
in $l_e$. This agrees very well with Kubo-Greenwood
calculations.\cite{Leconte2011a, Cresti2011}

\begin{figure}
  \centering
  \includegraphics[width=\columnwidth]{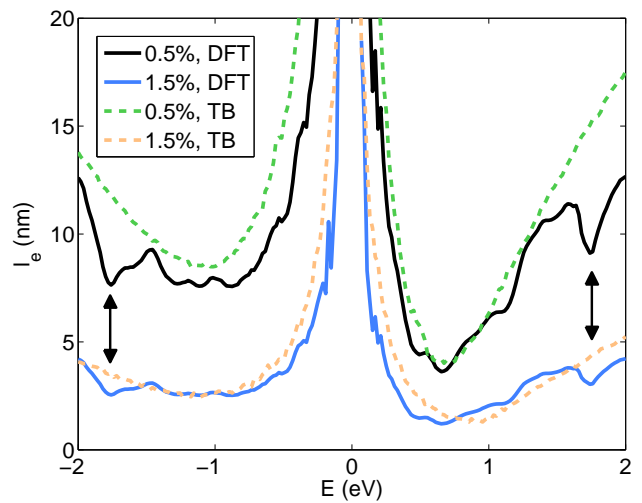}
  \caption{\label{fig:mfpo} (Color online) Mean free paths for
    epoxide adsorbates at two different concentrations.  DFT estimates
    based on single-defect systems are shown with solid lines, while
    ensemble-averaged TB results for large-scale many-defect systems
    are shown with dashed lines. The arrows indicate resonances not
    captured by the TB model.}
\end{figure}

The DFT estimates of $l_e$ are directly based on the scattering cross
section of Fig.~\ref{fig:epoxide} and the use of
Eq.~(\ref{eq:elastic}). To test the validity of this procedure, we
have performed large-scale TB simulations of systems with defect
concentrations of 0.5 and 1.5 \% (these values correspond to the
dashed lines in the inset of Fig.~\ref{fig:epoxide}), by
ensemble-averaging the results over different configurations of
adsorbate locations. The calculations have been performed using a 12
nm $\times$ 32 nm unit cell and an ensemble size of 50. Results
obtained through this procedure are compared with DFT predictions in
Fig.~\ref{fig:mfpo}.  In general, the results agree fairly well,
indicating that the transport properties of a large-scale system with
a moderate concentration of adsorbates can be reliably estimated based
on only a single-defect calculation. No intermediate steps are needed
for this procedure, and DFT can thus be directly used to calculate
various transport-related properties.

\begin{figure}
  \centering
  \includegraphics[width=\columnwidth]{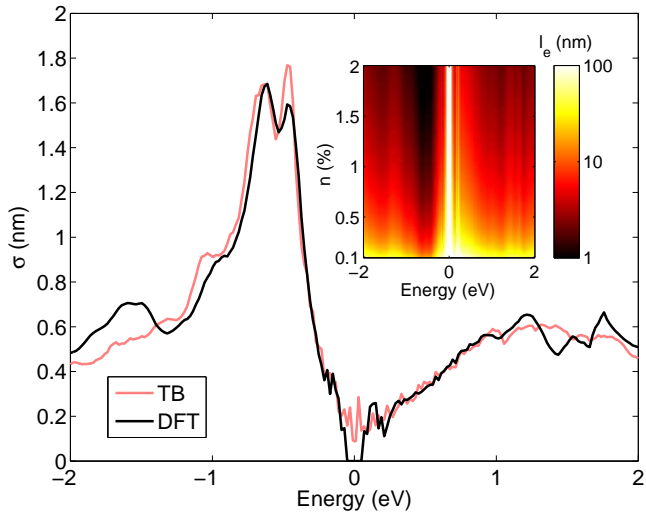}
  \caption{\label{fig:c_ga} (Color online) Scattering cross section for a carbon
    adatom on bulk graphene. The inset shows the elastic mean free path
    as a function of adatom concentration. The scale of the inset has been cut at 1 nm and 100 nm.}
\end{figure}

Carbon binds to graphene, like oxygen, on the carbon-carbon
bridge.\cite{Lehtinen2004,Nordlund1996} The electronic properties of
the carbon adsorbate are quite different from those of the epoxide
adsorbate, which is reflected in the clearly different scattering
cross sections.  Indeed, Fig.~\ref{fig:c_ga} shows that the
characteristic scattering fingerprint in the bulk scattering cross
section mirrors intriguingly the epoxide case with a peak below the
Fermi level.  This feature is a direct result of the $sp^3$ bond below
the Fermi level.  We are not aware of previous TB parameters for the
carbon adsorbate, so we have optimized the same set of parameters as
used above for epoxide\cite{Leconte2010} by matching the scattering
cross sections. The optimized parameters, using the same notation as
in the oxygen case, are $e_1=-4.00$ eV, $e_x=0.82$ eV, $e_z=3.85$ eV,
$t_1=-0.13 \, t$, $t_z=1.87 \, t$, and $t_x=1.30 \, t$.  Here we again
use $t=-2.6$ eV.  As can be seen in Fig.~\ref{fig:c_ga}, by using
these parameters the TB result matches very well with the DFT-based
cross section, apart from minor differences at large energies. As in
the case of epoxide, the elastic mean free path shows a clear
asymmetry between positive and negative energies. Furthermore, it is
important to note that both bridge adsorbates scatter charge carriers
mainly at finite energies, whereas the atop adsorbates cause very
strong scattering close to the Dirac point.

\section{Summary and conclusions}

The scaling approach using defect-specific energy dependent scattering
cross sections enables the modeling of electronic transport properties
of device-scale graphene systems with homogeneous defect
distributions.  The idea is that the scattering cross section can be
determined on the basis of an electron transport calculation for a
finite system containing a single defect.  In this work, we have
studied the sensitivity of determining the cross section on the
different electronic structure methods (DFT and TB) and different
system geometries (bulk graphene and graphene nanoribbons).  
We have found that a carefully constructed TB model works well 
for various different adsorbates. Furthermore, the scattering
cross section is an useful quantity in validating the TB model, and can
even be used to optimize the model parameters, as we have shown above
for the case of the bridge-carbon.

The scattering cross section provides a fingerprint which can be used
to categorize and compare different defects, especially adsorbate
species. An adsorbed hydrogen causes a strong peak in the cross
section at the Fermi level and the hydroxyl group exhibits also less
intense but clear peaks below and above the Fermi level.  Both the
hydrogen and hydroxyl adsorb on top of a carbon atom.  Oxygen in the
form of epoxide and carbon adatoms bind to two carbon atoms at the
bridge position.  In the case of epoxide the scattering cross section
indicates that electrons are scattered more readily than holes, while
for carbon adatoms the opposite is true.

\begin{acknowledgments}
  This research has been supported by the Academy of Finland through
  its Centers of Excellence Program (project no. 251748).
  Computational resources were provided by CSC--IT Center for Science
  Ltd and the Aalto Science-IT project.
\end{acknowledgments}

%

\end{document}